\documentclass[a4paper,12pt]{article}
\newcommand{\be}{\begin{equation}}
\newcommand{\ee}{\end{equation}}
\newcommand{\bea}{\begin{eqnarray}}
\newcommand{\eea}{\end{eqnarray}}
\newcommand{\ena}{\end{eqnarray}}

\def\half{\frac{1}{2}}
\usepackage{graphicx,amssymb,bm,latexsym}

\begin{document}
\topmargin 0pt \oddsidemargin 0mm
\renewcommand{\thefootnote}{\fnsymbol{footnote}}
\begin{titlepage}
\begin{flushright}
yymm.nnnn[hep-th]
\end{flushright}
\vspace{5mm}
\begin{center}
{\Large \bf D3-branes at angle in a linear dilaton pp-wave background}
\vskip 10mm
{Pratap K. Swain$^a$\footnote{E-mail: pratap@phy.iitkgp.ernet.in}}

\vskip 10mm
{\it $^a$ Department of Physics and Meteorology\\
Indian Institute of Technology Kharagpur, Kharagpur 721302, India}

\end{center}
\vspace{10mm} \centerline{{\bf{Abstract}}} \vspace{5mm}
We present a class of low energy type IIB supergravity solutions of two 
D3-branes rotated by an angle $\alpha \in SU(2)$ in a linear dilaton 
pp-wave background. The D-branes bound state configurations is 
found to preserve 1/16 of the total type IIB supersymmetries. We also
present a class of D1-D5 solution by applying T-duality on $D3\perp D3$
configuration.  
\end{titlepage}

\tableofcontents
\section{Introduction}
The study of string theory in time dependent background is one of the most interesting and challenging subjects till date. Untill now, it has not been fully explored. Due to their applications in solving puzzles of early universe cosmology, physicists  have taken much interest in finding time dependent supersymmetric solutions of supergravitities in ten and eleven dimensional space time \cite{{Liu:2002kb},{Hashimoto:2002nr},{Simon:2002cf},{Cai:2002sv},{Ohta:2003zh}}. In this connection, it has been proposed that a null linear dilaton background could play a toy model in solving the puzzles of early universe cosmology \cite{Craps:2005wd}. This background preserves 1/2 of the space-time supersymmetry, thereby, giving some hope of studying string theory in supersymmetric time dependent background. It has been shown that the space-like singularity can be resolved by proposing a kind of matrix string theory. At early time, the theory is strongly coupled, and hence the matrix degrees of freedom describe the correct physics. At late times, however, the theory can be seen as a free theory
\footnote{see \cite{null-dilaton} for discussions on various issues in
null dilaton background}. 
The AdS/CFT duality has also been examined in time dependent back ground 
with linear dilaton. It has been shown that the dual field theory can be 
realized with a time dependent Yang-Mills coupling constant \cite{Chu:2006pa}.

The study of D-branes is very useful in various developments of string theory. As it explains nonperturbative phenomena, it is very important to explore 
them further. In this context, supersymmetric D-brane solutions has been discussed in light-like linear dilaton background in literature \cite{{Chu:2006pa},{Das:2006dz}, {Nayak:2006dm},{Nayak:2006jc},{Ohta:2006sw},
{Maeda:2009tq},{Maeda:2010yk}}
. Motivated by the recent development in time dependent supersymmetric background and its necessity to explain cosmology, we present a class of time dependent linear dilaton pp-wave background solutions with various flux for two D3-branes rotated by an angle $\alpha \in SU(2)$ and its supersymmetric variation. In the
present case, the harmonic function depends both on tranverse coordinates and on the light cone time like coordinate $(x^+)$. Supergravity solutions 
for D-branes at angle has been constructed in various backgrounds earlier\cite{{Ohta:2006sw},{Maeda:2009tq},{Maeda:2010yk},{Berkooz:1996km},{Breckenridge:1997ar},{Hambli:1997uq},{Nayak:2002ty},{Nayak:2003hd}}.  

The outline of this paper is as follows. In section-2 we present the classical supergravity solutions for various configuratoins of D3-branes. First we
 construct the type IIB supergravity solutions for a single rotated D3-brane in a time dependent linear dilaton pp-wave background. Then we present 
solutions for two orthogonally interseting D3-branes and applying T-dualities, 
we find solutions for intersecting D1-D5 branes orthogonal to each other. 
Then we generalize it for two D3-branes rotated by an arbitrary angle $\alpha \in SU(2)$  to each other. In section-3, we study the supersymmetric properties
of the D3-branes at angle by solving the dilatino and gravitino variations
explicitly. We find, it preserves 1/16 of the total type IIB supersymmetries. 
In section-4, we conclude with some remarks.  

\section{D-brane Solutions}
In this section we present classical solutions of D3-branes
oriented along each other in the following linear dilaton pp-wave
background supplimented by the Neveu-Schwarz - Neveu-Schwarz(NS-NS) 
flux:
\begin{eqnarray}
ds^2  &=&  -2{dx^+}{dx^-} -{\mu}^2(x^+)\sum_{i=1}^4{x^2_i~{(dx^+)^2}}
+\sum_{a=5}^8{(dy^a)^2} + \sum_{i=1}^4{(dx^i)^2},
\cr & \cr e^{2\phi} &=& e^{-2f(x^+)},~~~~ H_{+12} = H_{+34} = 2\mu(x^+).
\nonumber \\ \label{pp}
\end{eqnarray}
\subsection{Single rotated D3-brane}
We would now like to write down the classical solutions for a single D3-brane rotated by a certain angle $\alpha \in SU(2)$ in the above linear dilaton pp wave background. The classical solutions for the D3-brane lying along $x^+, x^-, x^6$ \& $x^8$ directions in given by the following form of metric $(g_{\mu\nu})$, 
dilaton $(\phi)$, R-R fields $(F_{\mu\nu\rho\alpha\beta})$ and the NS-NS field
$(H_{\mu\nu\rho})$:
\begin{eqnarray}
ds^2 &=& {1 \over{\sqrt{1+\tilde{X_1}}}}\bigg( 2{d x^+}{d x^-} -{\mu^2}(x^+)\sum_{i=1}^4 {x^2_i~{(d x^+)^2}} 
\cr & \cr &+& [1+\tilde{X_1}\cos^2\alpha][(d x^5)^2 + (d x^7)^2] + [1+\tilde{X_1}\sin^2\alpha][(d x^6)^2 + (d x^8)^2]
\cr & \cr &+& 2\tilde{X_1}\cos \alpha \sin \alpha ({d x^7}{dx^8} - {dx^5}{dx^6})\bigg) +\sqrt{1 + \tilde{X_1}}\sum_{i=1}^4 (d x^i)^2,
\cr & \cr H_{+12} &=& H_{+34}= 2 \mu (x^+), 
\cr & \cr F^{(5)}_{+-68i} &=& -
e^{2f} {\partial_i \tilde{X_1}\over (1 + \tilde{X_1})^2}\cos^2\alpha, \cr & \cr
F^{(5)}_{+-67i} &=& e^{2f} {\partial_i \tilde{X_1}\over (1 + \tilde{X_1})^2}
\cos\alpha \sin\alpha, ~~~~~~ F^{(5)}_{+-57i} = e^{2f} {\partial_i
\tilde{X_1}\over (1 + \tilde{X_1})^2}\sin^2\alpha, \cr & \cr F^{(5)}_{+-58i} &=& -
e^{2f} {\partial_i \tilde{X_1}\over (1 + \tilde{X_1})^2} \cos\alpha \sin\alpha,
~~~~~ e^{2\Phi} = e^{-2f(x^+)}, ~~~ \tilde{X_1}(\vec r) = { 1\over 2}e^{-f(x^+)} \bigg({\ell_1 \over \vert \vec r - \vec r_1\vert}\bigg)^2. \nonumber
\\ \label{d3-nd-1}
\end{eqnarray}
where $(1+ \tilde{X_1})$ is the harmonic function in the transverse space of the D3-brane. Here the metric functions depend on both transverse space coordinates and a light-cone coordinate $x^+$. In order to explain the procedure for getting the above solutions from a D3-brane in a linear dilaton background, we start with the single unrotated D3-brane lying along $x^+, x^-,x^6$ and $x^8$ directions. By applying a rotation between $(x^5-x^6)$ and $(x^7-x^8)$ planes, with the rotation angle $(\alpha,\beta) = (0,\alpha)$ as described by \cite{Breckenridge:1997ar}, we get the configuration where the D3-brane is now oriented by an angle $\alpha \in SU(2)$ with respect to the original plane. For $\alpha = 0$, we get back the D3-brane solutions as given in \cite{Chu:2006pa} and for $f(x^+) = 0$, we get a tilted D3-brane solutions in pp wave background. The above solutions (\ref{d3-nd-1}) gives the generalization of the time independent tilted D-brane solutions to that of a time dependent one, with a linear dilaton pp-wave background. We have verified that the solutions (\ref{d3-nd-1}) solves all the type IIB field equations.

\subsection{Orthogonal Intesecton of two D3-branes}
Now we present the classical solutions of a system of two D3-branes lying relatively perpendicular to each other in a time dependent linear dilaton pp-wave background. The supergravity solutions of such system is given by
\begin{eqnarray}
ds^2 &=& (1+\tilde{X_1})^{-\half}(1+\tilde{X_2})^{-\half}\bigg[ 2{d x^+}{d x^-} -{\mu^2}(x^+)\sum_{i=1}^4 {x^2_i~{(d x^+)^2}}\bigg]\cr & \cr &+&(1+\tilde{X_1})^{-\half}(1+\tilde{X_2})^{\half} \bigg[(d x^5)^2 + (d x^7)^2\bigg] +(1+\tilde{X_1})^{\half}(1+\tilde{X_2})^{-\half}\bigg[(d x^6)^2 + (dx^8)^2\bigg]\cr & \cr &+& (1+\tilde{X_1})^{\half}(1+\tilde{X_2})^{\half}\sum_{i=1}^4 (d x^i)^2, 
\cr & \cr H_{+12} &=& H_{+34} = 2 \mu (x^+),
\cr & \cr F^{(5)}_{+-68i} &=& e^{2f}{\partial_i \tilde{X_2}\over (1 + \tilde{X_2})^2},
\cr & \cr F^{(5)}_{+-57i} &=&
- e^{2f} {\partial_i \tilde{X_1}\over (1 + \tilde{X_1})^2},~~~~e^{2\Phi} = e^{-2f(x^+)}, ~~~ \tilde{X_{1,2}}
(\vec r) = { 1\over 2} e^{-f(x^+)}\bigg({\ell_{1,2} \over \vert \vec r - \vec r_{1,2}\vert}\bigg)^2. \nonumber
\\ \label{d3-nd-2}
\end{eqnarray}
Here, the dilaton is linear in  light cone time coordinate $x^+$.
In this case, we start from two D3-branes parallel to each other and are lying along $x^+,x^-,x^6$ and  $x^8 $ directions. Now, applying SU(2) rotation to the branes at an angle $ \pi/2$  along $(x^5 - x^6)$ and $(x^7 - x^8)$ directions as described earlier. Now, one of the branes will lie along $x^+,x^-,x^5$ and  $x^7 $ directions and the other will lie along $x^+,x^-,x^6$ and  $x^8$. At the same time, the branes are delocalized along $(x^6 - x^8)$ and $(x^5 - x^7)$ planes respectively. We have checked that the solutions given equation (\ref{d3-nd-2}) also solves all type IIB field equations.

It is interesting to get more intersecting D-brane solutions from the (\ref{d3-nd-2}) by applying T-dualities along any of its longitudinal and transverse direction. For example, we can get orthogonal D1-D5 branes by applying T-dualities along $x^6$ and $x^8$ directions \cite{{Bergshoeff:1995as}, {Breckenridge:1996tt}}. The supergravity solutions for orthogonally intersecting D1-D5 branes in time dependent linear dilaton pp-wave background is given by:
\begin{eqnarray}
ds^2 &=& (1+\tilde{X_1})^{-\half}(1+\tilde{X_2})^{-\half}\bigg[2{dx^+}{dx^-} - {\mu^2}(x^+)\sum_{i=1}^4 {x^2_i~{{dx^+}^2}}\bigg]\cr & \cr &+& (1+\tilde{X_1})^{-\half}(1+\tilde{X_2})^{\half}\bigg[(dx^5)^2 +(dx^6)^2 +(dx^7)^2 + (dx^8)^2\bigg]\cr & \cr &+& (1+\tilde{X_1})^{\half}(1+\tilde{X_2})^{\half}\sum_{i=1}^4 (d x^i)^2, \cr & \cr H_{+12} &=& H_{+34} = 2 \mu (x^+),\cr & \cr F^{(3)}_{+-i} &=& e^{2f}{\partial_i X_2\over (1 + X_2)^2},\cr & \cr F^{(7)}_{+-5678i} &=& -e^{2f} {\partial_i X_1\over (1 + X_1)^2},~~~~e^{2\Phi} = e^{-2f(x^+)}\Bigg\{\frac{1+\tilde{X_2}}{1+\tilde{X_1}}\Bigg\}.
\nonumber \\ \label{d3-nd-3}
\end{eqnarray}
In the equation (\ref{d3-nd-3}), for $\tilde{X_1} = 0$, we get solution for D1-brane lying along $x^+$ and $x^-$ directions and for $\tilde{X_2} = 0$, we get solutions for D5-brane lying along $x^+,x^-,x^5,x^6,x^7$ and $x^8$ directions. 

\subsection{Two D3-branes at angle}
Now, we generalize the above solutions for two D3-branes rotating by an angle $\alpha \in SU(2)$. It has been discussed in literature that SU(2) rotation of D-brane preserves certain supersymmetries \cite{Berkooz:1996km}. Now, we present the low energy classical supergravity solutions of a system of two D3-branes oriented by an angle $\alpha \in SU(2)$ with respect to each other. The supergravity solutions of such a system in linear null dilaton pp-wave background with NS-NS and R-R fluxes is given by: 

\begin{eqnarray}
ds^2 &=& {1 \over{\sqrt{1+\tilde{X}}}}\bigg( 2{d x^+}{d x^-} -{\mu^2}(x^+)\sum_{i=1}^4 {x^2_i~{(d x^+)^2}} 
\cr & \cr &+& (1+\tilde{X_2})\big[(d x^5)^2 + (d x^7)^2\big]
 + (d x^6)^2 + (d x^8)^2
\cr & \cr &+& \tilde{X_1}\left[( \cos \alpha d x^5 - \sin \alpha d x^6)^2
+ ( \cos \alpha d x^7 + \sin \alpha d x^8)^2\right]\bigg) +
\sqrt{1 + \tilde{X}}\sum_{i=1}^4 (d x^i)^2 ,\cr & \cr H_{+12} &=& H_{+34}
= 2 \mu (x^+), \cr & \cr F^{(5)}_{+-68i} &=&
e^{2f}{\partial_i\Big\{{{\tilde{X_2} + \tilde{X_1} \cos^2\alpha + \tilde{X_1} {X_2}
\sin^2\alpha} \over {(1 + \tilde{X})}}\Big\}},\cr & \cr F^{(5)}_{+-58i}
&=& - F^{(5)}_{+-67i} = e^{2f}{\partial_i\Big\{{{\tilde{X_1} \cos\alpha
\sin\alpha} \over {(1 + \tilde{X})}}\Big\}},\cr & \cr F^{(5)}_{+-57i} &=&
- e^{2f} {\partial_i\Big\{{{(\tilde{X_1}+ \tilde{X_1} \tilde{X_2}) \sin^2\alpha}
 \over {(1 + \tilde{X})}}\Big\}},~~~~e^{2\Phi} = e^{-2f(x^+)}.
\label{d3-nd-pp}
\end{eqnarray}
and $\tilde{X}$ is given by
\begin{equation}\label{d3-2a}
\tilde{X} =\, {\tilde{X_1} + \tilde{X_2} + \tilde{X_1} \tilde{X_2} \sin^2 \alpha}.
\end{equation}
Here, $\tilde{X_{1.2}}$ is same as defined in equation (\ref{d3-nd-2}). To start with, two D3-branes parallel to each other and lying along $x^+, x^-, x^6$ and $x^8$ directions. Now, by applying rotation of an angle $\alpha \in SU(2)$ like discussed in case of perpendicular case, one will get the solutions as given in equation (\ref{d3-nd-pp}). We have verified that the above solutions given in (\ref{d3-nd-pp}) solves all type IIB field equations. If we put $\alpha$ as $\pi/2$, we will return to the equation (\ref{d3-nd-2}) and for $\tilde{X_2} = 0$ one will get back the equation (\ref{d3-nd-1}). The solutions given in equation (\ref{d3-nd-pp}) reduces to that of \cite{Nayak:2002ty} for $f =0$. Hence, the solutions given by us is a more generalized one in a class of time dependent pp-wave background with a linear dilaton.

\section{Supersymmetry Analysis}
It has already been discussed in the literature \cite{Berkooz:1996km} that
a system of $D$-branes oriented at certain angle $\alpha \in
SU(N)$ subgroup of rotations, with respect to each other, preserve
certain amount of unbroken supersymmetries. We would like to
analyze the fate of the supersymmetry in a linear dilaton
background through a particular example discussed in this paper by
solving the type IIB supersymmetry variations explicitly. The
supersymmetry variation of dilatino and gravitino fields of type
IIB supergravity in ten dimension, in string frame, is given by
\cite{{Schwarz:1983qr},{Hassan:1999bv}}:
\begin{eqnarray}
\delta \lambda_{\pm} &=& {1\over2}(\Gamma^{\mu}\partial_{\mu}\Phi
\mp {1\over 12} \Gamma^{\mu \nu \rho}H_{\mu \nu
\rho})\epsilon_{\pm} + {1\over
  2}e^{\Phi}(\pm \Gamma^{M}F^{(1)}_{M} + {1\over 12} \Gamma^{\mu \nu
  \rho}F^{(3)}_{\mu \nu \rho})\epsilon_{\mp},
\label{dilatino}
\end{eqnarray}
\begin{eqnarray}
\delta {\Psi^{\pm}_{\mu}} &=& \Big[\partial_{\mu} + {1\over
4}(w_{\mu
  \hat a \hat b} \mp {1\over 2} H_{\mu \hat{a}
  \hat{b}})\Gamma^{\hat{a}\hat{b}}\Big]\epsilon_{\pm} \cr
& \cr &+& {1\over 8}e^{\Phi}\Big[\mp
\Gamma^{\lambda}F^{(1)}_{\lambda} - {1\over 3!} \Gamma^{\lambda
\nu \rho}F^{(3)}_{\lambda \nu \rho} \mp {1\over 2.5!}
\Gamma^{\lambda \nu \rho \alpha \beta}F^{(5)}_{\lambda \nu \rho
\alpha
  \beta}\Big]\Gamma_{\mu}\epsilon_{\mp},
\label{gravitino}
\end{eqnarray}

where we have used $(\mu, \nu, \rho, \lambda)$ to describe the ten dimensional space-time indices, and hat's represent the corresponding tangent space indices. Solving the above two equations for the solutions describing the system of two $D3$-branes as given in equation (\ref{d3-nd-pp}), we get several conditions on the spinors. First the vanishing dilatino variation gives:
\begin{equation}
\Gamma^{\hat +}~\epsilon_{\pm} = 0, \label{light-cone}
\end{equation}
and 
\begin{equation}
\left(1 - \Gamma^{\hat1\hat2\hat3\hat4}\right)~\epsilon_{\pm} = 0. 
\label{light-cone-1} 
\end{equation}

Gravitino variations gives the following conditions on the
spinors:
\begin{eqnarray}
\delta \psi_+^{\pm} &\equiv & \partial_{+}\epsilon_{\pm} +\Bigg[ \frac{1}{8} (1+\tilde{X})^{-1}\partial_+{\tilde{X}} + \frac{1}{8} (1+\tilde{X})^{-\frac{3}{2}}\Gamma^{\hat{-}\hat{i}}\partial_i{\tilde{X}} \Bigg] \epsilon_{\pm}\cr & \cr
&+& \Bigg[\frac{1}{4}(1+\tilde{X})^{-\frac{1}{2}}(1+ \tilde{X_1}\sin^2\alpha)^{-1}\sin\alpha\cos\alpha
\partial_+{\tilde{X_1}} (\Gamma^{\hat{7}\hat{8}} - \Gamma^{\hat{5}\hat{6}})\Bigg] \epsilon_{\pm}
\cr & \cr &{\mp}& \frac{1}{2}\frac{\mu}{\sqrt{1+\tilde{X}}}
(\Gamma^{\hat{1}\hat{2}} + \Gamma^{\hat{3}\hat{4}})\epsilon_{\pm} 
\cr & \cr &{\mp}& {e^{f/2}\over 8} ~\Gamma^{\hat +\hat -\hat 6\hat 8\hat i}~\Bigg[ {{(1 + \tilde{X_1} \sin^2 \alpha)^2 ~{\partial_i \tilde{X_2}} + \cos^2\alpha ~{\partial_i \tilde{X_1}}} \over {(1 + \tilde{X})^{3/2}(1 + \tilde{X_1} \sin^2 \alpha)}}\Bigg] \Gamma^{\hat -} \epsilon_{\mp} 
\cr &\cr &{\mp}& {e^{f/2}\over 8}  \Gamma^{\hat +\hat -\hat 5\hat 7\hat i}~\Bigg[ {1\over (1 + \tilde{X})^{5/2}}(1 + \tilde{X_1} \sin^2 \alpha)
\cr & \cr &\times & ({{\tilde{X_1}}^2 \cos^2\alpha \sin^2\alpha }~{\partial_i \tilde{X_2}} + (1 + \tilde{X_2})^2 \sin^2 \alpha ~{\partial_i \tilde{X_1}})
\cr & \cr &-& {{({\tilde{X_1}}^2 \cos^2\alpha \sin^2\alpha)(~(1 + \tilde{X_1} \sin^2 \alpha)^2~{\partial_i \tilde{X_2}} + \cos^2\alpha ~{\partial_i \tilde{X_1}})}\over (1 +\tilde{X})^{5/2} (1 + \tilde{X_1} \sin^2 \alpha)}
\cr & \cr &+& {1\over{(1 +\tilde{X})^{5/2}}} \bigg(2{\tilde{X_1}}^2 \cos^2\alpha \sin^2\alpha ~{\partial_i\tilde{X_2}} + 2{\tilde{X_1}}^3 \cos^2\alpha \sin^4\alpha ~{\partial_i \tilde{X_1}}
\cr &\cr &-& 2 \tilde{X_1}(1 + \tilde{X_2})\cos^2\alpha \sin^2\alpha ~{\partial_i\tilde{X_1}}\bigg) \Bigg]\Gamma^{\hat -} \epsilon_{\mp}
\cr & \cr &{\mp}&{e^{f/2}\over 8}~\bigg\{{{\Gamma^{\hat +\hat -\hat 5\hat 8\hat i}~-~\Gamma^{\hat +\hat -\hat 6\hat 7\hat i}} \over {(1 + \tilde{X})^2( 1 +\tilde{X_1} \sin^2\alpha)}}\bigg\} \Bigg[(- \tilde{X_1}\cos\alpha \sin \alpha
~{\partial_i \tilde{X_2}}
\cr & \cr &+& (1+\tilde{X_2})\sin\alpha \cos\alpha
~{\partial_i \tilde{X_1}} - {\tilde{X_1}}^2 \sin^3\alpha \cos\alpha~{\partial_i \tilde{X_1}})( 1 + \tilde{X_1} \sin^2\alpha)
\cr & \cr &+& \tilde{X_1} \cos\alpha \sin\alpha ( 1 + \tilde{X_1}
\sin^2\alpha)^2~{\partial_i \tilde{X_2}}
\cr & \cr &+& \tilde{X_2}\cos^3\alpha
\sin\alpha ~{\partial_i \tilde{X_1}}\Bigg]
\Gamma^{\hat -} \epsilon_{\mp} = 0. 
\label{gravitino-1}
\end{eqnarray}
\begin{eqnarray}
\delta {\psi_-}^{\pm} &\equiv & \partial_{-}\epsilon_{\pm}=0, ~~~~
\delta \psi_a^{\pm} \equiv \partial_{a}\epsilon_{\pm}=0 ~~~~~ (a=5,6,7,8).
\cr & \cr \delta {\psi_i}^{\pm} &\equiv &
\partial_{i}\epsilon_{\pm} + \frac{1}{8} \frac{\partial_i \tilde{X}}{(1+\tilde{X})}\epsilon_{\pm}=0
\label{gravitino-2}.
\end{eqnarray}

While writing down the $\delta{\psi_i^{\pm}}$ variation, we have made use of the
conditions (\ref{light-cone}), and (\ref{light-cone-1}).
Now, the variation $\delta {\psi_+}^{\pm}$ in equation (\ref{gravitino-1}) is
further solved by imposing the following conditions, including
(\ref{light-cone}) and (\ref{light-cone-1}).
\begin{eqnarray}
&& (\Gamma^{\hat 5\hat 8} -\Gamma^{\hat 6\hat 7})\epsilon_{\mp} =  0,~~~~(\Gamma^{\hat 5\hat 7} + \Gamma^{\hat 6\hat 8})\epsilon_{\mp} = 0,~~~~
\cr & \cr && (\Gamma^{\hat{5}\hat{6}} - \Gamma^{\hat{7}\hat{8}})
\epsilon_{\pm} = 0, \label{rotation}
\end{eqnarray}
and the usual D3-brane supersymmetry conditions of flat space:
\begin{eqnarray}
\Gamma^{\hat +\hat -\hat 6\hat 8}\epsilon_{\mp} =
\epsilon_{\pm},~~~~ \Gamma^{\hat +\hat -\hat 5\hat
7}\epsilon_{\mp} = \epsilon_{\pm}. \label{brane}
\end{eqnarray}
After imposing all these conditions (\ref{light-cone}), (\ref{light-cone-1}), (\ref{rotation}) and (\ref{brane}), the equation (\ref{gravitino-1}) and equation (\ref{gravitino-2}) reduce to the following form:

\bea
&&\partial_+ \epsilon_{\pm} + \frac{1}{8} \frac{\partial_+ \tilde{X}}{(1+\tilde{X})} \epsilon_{\pm} = 0 , \cr & \cr && \partial_{i}\epsilon_{\pm} + \frac{1}{8}
\frac{\partial_i \tilde{X}}{(1+\tilde{X})}\epsilon_{\pm}=0. 
\label{pp-1}
\ena 
The above two equations are clearly solved by the following spinor,
\bea 
\epsilon_{\pm} = (1+\tilde X)^{-{1/8}}\epsilon_{0} ,
\label{pp-2}
\ena 
where $\epsilon_0$ is a constant spinor. Let's now count the total number of 
independent conditions, we have taken for solving the dilatino and gravitino
variations. First imposing (\ref{light-cone}), the dilatino variation 
(\ref{dilatino}) is satisfied completely, thereby breaking 1/2 supersymmetry. 
Further to satisfy all the gravitino variations, we have made use of the
conditions (\ref{rotation}) and (\ref{brane}), in addition to 
(\ref{light-cone-1}), thereby breaking 1/8 of the remaining supersymmetry,
as these are only three independent conditions.
Hence, the system of two D3-branes rotated by an angle $\alpha \in SU(2)$ in 
a linear dilaton time dependent pp-wave background preserves 
1/16 of total type IIB spacetime supersymmetries.

\section{Conclusions}
In this paper we have constructed the classical solution of two D3-branes rotated by an  angle $\alpha \in SU(2)$ with respect to each other in a linear dilaton time dependent pp- wave background. The supersymmetry of this solution has been checked by solving the dilatino and gravitino variations, and we found, it preserves 1/16 supersymmetries of type IIB string theory. The open string 
construction for various branes and their bound states can be done by following
\cite{Narayan:2009pu,Madhu:2009jh} and chosing
the boundary conditions appropriately. We wish to come back to this issue
in future. 


\end{document}